\documentclass{aa}
\usepackage{graphicx}
\usepackage{txfonts}
\usepackage{natbib}
\usepackage{color}
\bibpunct{(}{)}{;}{a}{}{,}    
\newcommand{\beq}{\begin{equation}}
\newcommand{\eeq}{\end{equation}}
\newcommand{\bea}{\begin{eqnarray}}
\newcommand{\eea}{\end{eqnarray}}

\newcommand{\url}[1]{{\tt #1}}

\def\gapp{\lower 3pt\hbox{${\buildrel > \over \sim}$}\ }
\def\lapp{\lower 3pt\hbox{${\buildrel < \over \sim}$}\ }

\usepackage{color}
\usepackage{ulem}

\newlength{\linwx}
\begin{document}
\title{Stellar irradiated discs and implications on migration of embedded planets II: accreting discs}
\author{
Bertram Bitsch \inst{1},
Alessandro Morbidelli \inst{1},
Elena Lega \inst{1},
Aur\'{e}lien Crida \inst{1}
}
\offprints{B. Bitsch,\\ \email{bertram.bitsch@oca.eu}}
\institute{
University of Nice-Sophia Antipolis, CNRS, Observatoire de la C\^{o}te d'Azur,
Laboratoire LAGRANGE, BP4229, 06304 NICE cedex 4, FRANCE
}
\abstract
{ The strength and direction of the migration of embedded low mass planets depends on the disc's structure. It has been shown that, in discs where the viscous heating is balanced by radiative transport, the migration can be directed outwards, a process that extends the lifetime of growing planetary embryos.
 } 
{ In this paper we investigate the influence of a constant $\dot{M}$-flux through the disc, as well as the influence of the disc's metallicity on the disc's thermodynamics. We focus on $\dot{M}$ discs, which have a net mass flux through them. Utilizing the resulting disc structure, we determine the regions of outward migration in the disc.
} 
{ We performed numerical hydrosimulations of $\dot{M}$ discs with viscous heating, radiative cooling, and stellar irradiation in 2D in the r-z plane. We used the explicit/implicit hydrodynamical code FARGOCA that includes a full tensor viscosity and stellar irradiation, as well as a two-temperature solver that includes radiation transport in the flux-limited diffusion approximation. The migration of embedded planets is studied by using torque formulae.
 } 
{
 For a disc of gas surface density $\Sigma_G$ and viscosity $\nu$, we find that the disc's thermal structure depends on the product $\Sigma_G \nu$ and the amount of heavy elements, while the migration of planets, besides the mentioned quantities, depends on the amount of viscosity $\nu$ itself. As a result of this, the disc structure cannot be approximated by simple power laws. During the lifetime of the disc, the structure of the disc changes significantly in a non-linear way in the inner parts. In the late stages of the disc's evolution (characterized by low $\dot{M}$), outward migration is only possible if the metallicity of the disc is high. For low metallicity, planets would migrate inwards and could potentially be lost to the star. 
}
{
 The presented disc structures and migration maps have important consequences on the formation of planets, since they can give hints on the different formation mechanisms for different types of planets as a function of metallicity. 
}
\keywords{accretion discs -- planet formation -- hydrodynamics -- radiative transport -- planet disc interactions -- stellar irradiation}
\maketitle
\markboth
{Bitsch et al.: Stellar irradiated discs and implications on migration of embedded planets}
{Bitsch et al.: Stellar irradiated discs and implications on migration of embedded planets}

\section{Introduction}
\label{sec:introduction}
In recent years, it has been shown that the migration of low mass planets ($\approx 10-20 M_E$) can be changed from inwards to outwards in discs with heating and cooling \citep{2006A&A...459L..17P, 2008A&A...478..245P, 2008ApJ...672.1054B, 2008A&A...485..877P, 2008A&A...487L...9K, 2009A&A...506..971K, 2010MNRAS.408..876A}. The heating in these previous works is provided by viscous friction, while for the cooling, a local cooling rate (in 2D simulations) or a radiative diffusion (in 3D simulations) is utilized. This change of migration has important consequences for the formation of planets, as it would provide for a zero-migration radius in the discs, where planets can survive \citep{2010ApJ...715:L68,2011A&A...536A..77B}. Along with that, these points of convergent migration in the disc can trap planetary cores in resonances \citep{2013A&A...553L...2C}. But these cores could then break free from the resonance (due to turbulence), then collide and form the cores of giant planets \citep{2013A&A...558A.105P}.

Only recently have \citet{2013A&A...549A.124B} (hereafter Paper I) shown the importance of stellar irradiation on the disc structure and the migration of low mass planets in numerical simulations. In Paper I, we did show that the change of the disc structure from a shadowed to a flaring disc greatly influences the migration of embedded bodies. Roughly speaking, a local increase in the aspect ratio $H/r$ of the disc results in inward migration, while a local decrease in $H/r$ leads to outward migration. As a result, planets can migrate outwards in shadowed regions of the disc.

In Paper I, we pointed out that the opacity in the disc plays an important role determining the disc structure, since the opacity is relevant for the absorption of stellar irradiation and for the cooling of the disc. We only considered equilibrium discs that have no net-mass flux through the disc. In this work, we want to expand to discs with a constant mass flow $\dot{M}$ through the disc. In principle a viscous accretion disc is only accreting in the inner parts of the disc, while it is viscously spreading in the outer parts of the disc \citep{1974MNRAS.168..603L} and actually behaves as a decretion disc there. If $\dot{M}$ strongly depends on $r$, the surface density profile varies, so that the high $\dot{M}$ regions empty, and the low $\dot{M}$ regions fill, which smooths the variations in $\dot{M}$ on a quicker timescale for sharper variations. Thus, it is reasonable to assume that $\dot{M}$ is almost independent of $r$ at a given moment in the inner parts of the disc, on which we focus here. This accretion rate $\dot{M}$ can actually be measured in observable discs.


By studying different $\dot{M}$ rates for discs, we explore the disc structures as a function of disc evolution, where $\dot{M}$ decreases with time. Additionally, we study the influence of metallicity on the disc structure, because the dust-to-gas ratio can change in time as the disc evolves. The disc's dust-to-gas ratio could decrease if the radial drift gets rid of the dust particles faster than the gas falls onto the star or if planetesimals form efficiently, binding the dust. But it could increase as the gas is removed, and dust is left behind or is regenerated (e.g. due to collisional grinding); in the extreme case, this leads to debris discs, where the dust-to-gas ratio is infinite. 

Additionally, we investigate the influence of the disc structure on the migration of embedded planets. Since 3D simulations of planets are computationally very expensive, we relate back to using torque formulae for prescribing the expected migration of planets in the disc. We therefore are able to lay out a complete history of migration during the evolution of protoplanetary discs, which would not be possible in 3D.

The paper is structured as follows. First, we give an overview over the numerical methods used in this work in section~\ref{sec:methods}. We then discuss about the disc structure and planetary migration in discs that have a metallicity of $0.01$ in section~\ref{sec:Mdotdiscconst}. The implications of different rates of metallicity on the disc structure and migration is described in section~\ref{sec:Mdotdiscvary}. We then point out what roles viscosity and gas surface density play for the migration of planets in discs with constant $\dot{M}$ rates in section~\ref{sec:InfluenceSnu}. In section~\ref{sec:discussion}, we discuss the implications of our results on the formation and migration of planets in accretion discs. Finally, we summarize in section~\ref{sec:summary}.

\section{Methods}
\label{sec:methods}

The protoplanetary disc is treated in this study as a three-dimensional (3D), non-self-gravitating gas whose motion is described by the Navier-Stokes equations. Without any perturber like planets, the disc is an axisymmetric structure. We therefore can use only one grid cell in azimuthal direction, making the computational problem de-facto 2D in the radial and vertical ($r-z$ plane) directions, where we utilize spherical coordinates ($r$-$\theta$). The colatitude $\theta$ is measured in such a way that $\theta=90^\circ$ is the midplane of the disc. A clear picture of the source of turbulence inside accretion discs is under strong debate (see e.g. \citet{Turner2014}). Nevertheless, some sort of viscosity is needed to drive the accretion disc, so we treat the viscosity with an $\alpha$-prescription \citep{1973A&A....24..337S}. The dissipative effects can then be described via the standard viscous stress-tensor approach \citep[e.g.][]{1984frh..book.....M}. We also include the irradiation from the central star, which was described in detail in Paper I. For that purpose we modified and substantially extended an existing multi-dimensional hydrodynamical code FARGOCA, as presented in \citet{Lega2013}.

The radiative energy associated with viscous heating and stellar irradiation is then diffused through the disc and emitted from its surfaces. To describe this process we utilize the flux-limited diffusion approximation \citep[FLD,][]{1981ApJ...248..321L}, an approximation that allows the transition from the optically thick mid-plane to the thin regions near the disc's surface.

The hydrodynamical equations solved in the code have already been described in detail \citep{2009A&A...506..971K}, and the two-temperature approach for the stellar irradiation was described in detail in Paper I, so we refrain from quoting it here again. As in Paper I, we take $R_\star = 3 R_\odot$ and $T_\star = 5600$K, which gives the flux from the star:
\begin{equation}
 F_\star = \frac{R_\star^2 \sigma T_\star^4}{r^2} \ ,
\end{equation}
where $\sigma$ is the Stefan–Boltzmann constant and $r$ the distance to the star. Stellar heating is responsible for keeping the disc flared in the outer parts of the disc (Paper I).

\subsection{Opacity}

The opacity is a crucial parameter when using an energy equation with stellar heating. The Planck mean opacity $\kappa_P$ is used in the coupled two-energy equations (radiative energy and thermal energy), the Rosseland mean opacity $\kappa_R$ is used for the radiative cooling, and the stellar opacity $\kappa_\star$ is used for the absorption of stellar photons in the upper layers of the disc. For more information on how the different opacities are connected to the energy equation, see Paper I.

\begin{figure}
 \centering
 \includegraphics[width=1.0\linwx]{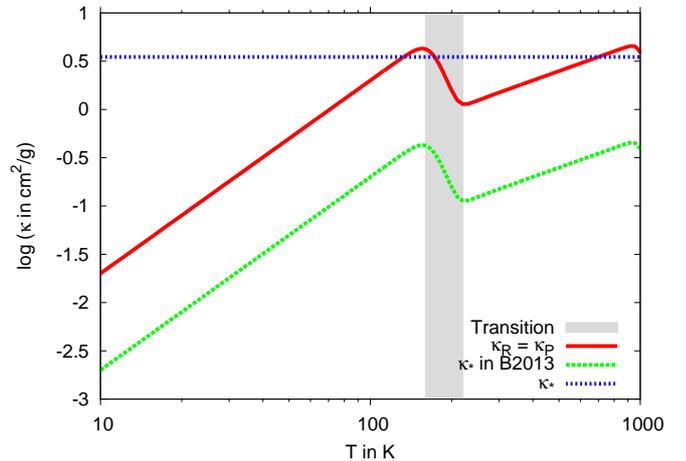}
 \caption{Opacity profiles of the opacities used in the code. The different bumps in the profile correspond to opacity transition, where the ice line is located at $\approx 170$K. The grey area marks the region where the opacity changes due to the melting of ice grains. The maximum temperature during the simulation is reached in the inner parts of the disc, where viscous heating is dominant and can reach up to $\approx 900$K for the $1 \times 10^{-7} M_\odot/yr$ case. Opacities are plotted for a metallicity of $0.01$.
   \label{fig:kappa}
   }
\end{figure}

In Fig.~\ref{fig:kappa}, the three different opacities used in the code are displayed. For the Rosseland mean opacity $\kappa_R$ and for the Planck mean opacity $\kappa_P$, we use the same opacity law, since these opacities do not differ that much in the temperature region we are interested in (Paper I). However, compared to our previous work in Paper I, we changed the stellar opacity $\kappa_\star$. In Paper I, we took the stellar opacity to be $0.1$ of the Rosseland mean opacity, as indicated in Fig.~\ref{fig:kappa}. However, the stellar opacity depends on the temperature of the star $T_\star$ and not on the temperature of the gas, since dust grains absorb photons regardless of their temperature. As a consequence, $\kappa_\star$ is shown as a constant line in Fig.~\ref{fig:kappa}. The value of $\kappa_\star = 3.5$cm$^2$/g corresponds to the typical value in a disc with a mixture of ice and silicate grains for $T_\star =5600$K. The change in $\kappa_\star$ compared to Paper I will lead to a higher level of absorption of stellar photons in the upper layers of the disc, because the optical depth $\tau = \kappa_\star \rho \Delta r$ is much larger. Eventually, since the opacity was very low in the upper regions of the disc in the Paper I study, this will lead to flared discs for much lower gas densities than in Paper I. But at the same time, the main arguments about the disc structure (flared vs. non-flared) and for the migration still hold in the sense of the Paper I. 

In the following, we define metallicity as the ratio of heavy elements to gas. In the condensed form, as ice or silicate grains, these heavy elements are, in our work, only $\mu m$ in size. We define $\Sigma_Z$ as the surface density of heavy elements in vapour or micrometer grain form and $\Sigma_G$ as the gas surface density. Thus, the metallicity $Z$ is simply the ratio $Z=\Sigma_Z / \Sigma_G$, assumed to be independent of $r$ in the disc. This means that if grain growth occurs and the total amount of heavy elements (independent of size) stays the same, then the metallicity in our sense is still reduced.

\subsection{Surface density}
\label{subsec:surfdens}

In a steady state accretion disc, the inward velocity of the gas is
\begin{equation}
 v_r = - \frac{3 \nu}{2 r} \ ,
\end{equation}
where $\nu=\alpha H^2 \Omega_K$ is the $\alpha$-viscosity \citep{1973A&A....24..337S} with $H$ the height of the disc, $\Omega_K$ the Keplerian orbital frequency, and $r$ denotes the orbital distance. With the radial velocity we can define an accretion rate $\dot{M}$ as
\begin{equation}
 \dot{M} = - 2\pi r \Sigma_G v_r = 3 \pi \nu \Sigma_G \ ,
\end{equation}
where $\Sigma_G$ is the gas surface density. The slope of the surface density can be calculated using the viscosity 
\begin{equation}
\label{eq:discvisc}
 \nu = \alpha H^2 \Omega \propto \alpha h_0^2 r^{2a} r^{-3/2} \,
\end{equation}
with $h=H/r$ and where $a=9/7$ as $H/r \propto r^{2/7}$ (\citet{1997ApJ...490..368C}, Paper I). The proportionality, $\propto r^{2/7}$ describes the flaring index of the disc, which gives for a steady state accretion disc, where $\dot{M}$ is constant for all radii ($(d/dr) \ \dot{M} = 0$),
\begin{equation}
 \dot{M} \propto h_0^2 r^{2a} r^{-3/2} \Sigma_{G,0} r^{-s_{\dot{M}}} \ ,
\end{equation}
where $s_{\dot{M}}$ denotes the power law index of the gas surface density in the $\dot{M}$ disc. Here, $\Sigma_{G,0}$ is the value of the gas surface density at $r_0=1$AU. This leads to
\begin{equation}
 2a-3/2-s_{\dot{M}} = 0 \quad \Leftrightarrow \quad 18/7 - 3/2 = s_{\dot{M}} \quad \Leftrightarrow \quad s_{\dot{M}} = 15/14 \ .
\end{equation}
We use this slope of surface density for the initial conditions in all our simulations. Vertically, we initially set the density $\rho$ of the disc to be in hydrostatical equilibrium
\begin{equation}
 \rho (r,\theta) = \rho_0 (r) \, \exp \left[ - \frac{(r \cos \theta)^2}{2 H^2} \right] \ ,
\end{equation}
where $\rho_0 (r)$ is the initial density.

For all simulations an adiabatic index of $\gamma=1.4$ and a mean molecular weight of $\mu = 2.3$ is set.

\subsection{Boundary conditions}
\label{subsec:boundary}

To get a disc with a given $\dot{M}$ rate, we impose an $\dot{M}$-rate at the outer boundary, while we leave the inner boundary open. This way, the density structure inside the disc can change from the initial conditions and form bumps in the surface density that could not be created if an $\dot{M}$ boundary was applied at the inner boundary as well, because the total mass inside the $\dot{M}$ disc would stay constant and no refilling could take place. The important quantities that have to be treated in a special way at the boundaries are the radial velocity $v_r$ and the gas surface density $\Sigma_G$ in the ghost rings. The details for the boundary conditions can be found in Appendix~\ref{ap:boundary}.

\subsection{Disc evolution}
\label{subsec:evolution}

\begin{figure}
 \centering
 \includegraphics[width=1.0\linwx]{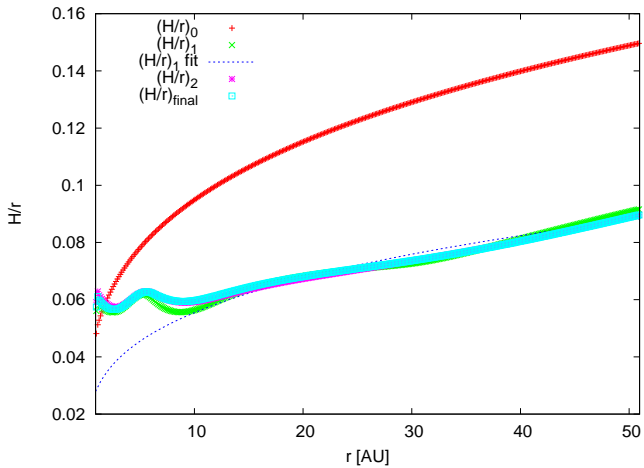}
 \caption{Evolution of $H/r$ in the disc for the different stages. The disc features $\dot{M} = 2 \times 10^{-7} M_\odot / yr$. At the beginning, in the $(H/r)_0$ stage, $\alpha_0 = 0.001$, which we change to $\alpha_1 = 0.003$ after the stage $(H/r)_1$ is reached. The stage $(H/r)_2$ shows the aspect ratio after the viscosity has change and $(H/r)_{final}$ shows the final stage after the surface density has been re-arranged.
   \label{fig:Hrevolve}
   }
\end{figure}

The initial conditions provided in section~\ref{subsec:surfdens} give an initial aspect ratio of the disc, $(H/r)_0$, which has to be set to a higher value as the equilibrium $H/r$ in order to get a flared disc, because a non flared disc can not become flared any more \citep{2002A&A...395..853D}, see Fig.~\ref{fig:Hrevolve}. In time the viscosity would then drop (as $H$ drops and $\nu = \alpha H^2 \Omega_K$), until an equilibrium state is reached. Starting a simulation like that would imply that $H$, therefore $\Sigma$, change a lot in time to guarantee a constant $\dot{M}$. If $H$ drops by a factor of $2-3$ as observed in Paper I, the surface density therefore has to rise by a factor of $4-9$. The time for the disc to adjust to that would be extremely long, since the disc has to be refilled from the outer boundary on a viscous timescale, which is essentially the lifetime of the disc.

We impose several steps in the simulations to save computation time. We first keep the viscosity constant in time and equal to its initial value $\nu_0=\alpha h_0^2 r^{15/14}$, by using the initial $h_0=(H/r)_0$ configuration to determine the viscosity during the whole first step. As the disc evolves, it compresses towards midplane and we reach an equilibrium state with a flared disc profile in the outer parts of the disc. We then fit this profile, which is labelled $(H/r)_1$ in Fig.~\ref{fig:Hrevolve}, with a new $H/r$ value, where $h_1 \propto r^{2/7}$ (shown as $(H/r)_{1,fit}$ in Fig.~\ref{fig:Hrevolve}). With this new estimate of $H/r$, we can compute a new $\alpha$ value for the viscosity by comparing it to $(H/r)_0$. The change of $H$ in the viscosity is compensated by the change of $\alpha$, so that the disc will have the same viscosity as in the first step and therefore the $\dot{M}$ rate is conserved as well.

The simulation is then restarted with the new $\alpha$ value (which is given by $\alpha_{end} = \alpha_0 h_0^2 / h_1^2$), and the viscosity $\nu$ is now given by  eq.~\ref{eq:discvisc}, with the local value of $H$ to account for bumps and dips that might eventually form in the disc. For the outer parts, these changes of $\nu$ are not important because the disc structure is determined by stellar irradiation and not by viscous heating. In the inner parts, where viscous heating is the dominant heat source, the viscosity changes compared to the $(H/r)_0$ case, leading to change in the disc structure. However, these changes are minimal. The advantage of this procedure is that the $\dot{M}$ rate is the same as in the initial simulation with no need to refill $\Sigma_G$ as $H$ drops.

We then reach a new equilibrium state $(H/r)_2$, where the inner parts of the disc have changed compared to $(H/r)_1$, because the viscosity changed. But, since the disc still needs to re-arrange the surface density on a viscous timescale, we relate back to a locally isothermal simulation, where we plug in the disc structure and sound speed as shown in the $(H/r)_2$ disc. The disc is then evolved until a time independent mass flux through the disc is achieved. After that, we go back to the fully radiative energy equation with stellar irradiation to account for changes in the heating due to the re-arranged gas surface density $\Sigma_G$ and arrive at $(H/r)_{final}$, which gives the final state of the disc. Please note that this whole process can only work as the outer disc is supported by stellar irradiation, keeping the input of mass at the outer boundary constant, not only in time, but also in vertical spacing because $H$ stays constant.

In this way, much computation time is saved, because the computation time is mainly dominated by the stellar-irradiation algorithm that can take up to $90\%$ of the whole computation time. Especially in the state where the surface-density has to be re-arranged, due to changes in viscosity this makes a big difference.

In principle with time the metallicity of small dust grains in the disc can change (e.g. due to grow of planetesimals), which would change the disc structure (see section~\ref{sec:Mdotdiscvary}). This would imply that disc needs to re-arrange the gas inside, which happens on a viscous timescale. However, as we show below, this re-arranging of mass is much less pronounced in the outer parts of the disc, where the viscous timescale is much longer compared to the inner parts ($\tau_{visc}=r^2/\nu = r^2 / (\alpha H^2 \Omega_K)$). This is because the change in viscosity in the outer parts is much smaller than the change in the inner parts of the disc, because the $H/r$ profile stays constant as it is dominated by stellar irradiation. This indicates that the steady state of constant $\dot{M}$ can be achieved in the inner parts on a timescale that is shorter than the lifetime of the disc, which is determined by the viscous timescale of the outer parts of the disc.

\subsection{Numerical setup}
\label{subsec:numsetup}

We performed several simulations for different $\dot{M}$ rates corresponding to different stages of the discs' lifetime that are listed in Table~\ref{tab:Mdotmet1}. In the first sets of simulations, we assumed that the dust is bound perfectly to the gas and as the gas accretes onto the star, the dust decreases at the same rate, keeping the dust-to-gas ratio constant. In the second set of simulations, we assumed that the dust is not perfectly bound to the gas at the later stages of the disc's evolution, resulting in an increasing and decreasing dust-to-gas ratio, depending on the favoured scenario for the evolution of dust particles. 

All simulations were performed on a grid with $386 \times 66$ active cells in $r-\theta$ direction, where the opening angle of the numerical simulation is $20^\circ$, meaning that $70^\circ < \theta < 90^\circ$. The radial domain extends from $1$AU to $50$AU.

{%
\newcommand{\mc}[3]{\multicolumn{#1}{#2}{#3}}
\begin{table}[b]
 \centering
 \begin{tabular}{|l|l|l|l|}\cline{1-4}
 \textbf{$\dot{M}$-rate [$M_\odot / yr$]} & \textbf{Metallicity} & \textbf{$\Sigma_{G,0}$} & $\alpha_{end}$ \\\hline
 \mc{1}{|l|}{$2 \times 10^{-7}$} & 0.01 & $5.385 \times 10^{-3}$ & 0.003  \\\hline
 \mc{1}{|l|}{$1 \times 10^{-7}$} & 0.01 & $2.693 \times 10^{-3}$ & 0.003  \\\hline
 \mc{1}{|l|}{$5 \times 10^{-8}$} & 0.01 & $1.346 \times 10^{-3}$ & 0.003  \\\hline
 \mc{1}{|l|}{$1 \times 10^{-8}$} & 0.01 & $2.693 \times 10^{-4}$ & 0.003  \\\hline
 \mc{1}{|l|}{$5 \times 10^{-9}$} & 0.01 & $1.346 \times 10^{-4}$ & 0.003  \\\hline \hline \hline
 \mc{1}{|l|}{$5 \times 10^{-8}$} & 0.02 & $1.346 \times 10^{-3}$ & 0.003  \\\hline
 \mc{1}{|l|}{$1 \times 10^{-8}$} & 0.03 & $2.693 \times 10^{-4}$ & 0.0026  \\\hline
 \mc{1}{|l|}{$5 \times 10^{-9}$} & 0.05 & $1.346 \times 10^{-4}$ & 0.0026  \\\hline
 \mc{1}{|l|}{$5 \times 10^{-8}$} & 0.001 & $1.346 \times 10^{-3}$ & 0.0026  \\\hline
 \mc{1}{|l|}{$1 \times 10^{-8}$} & 0.001 & $2.693 \times 10^{-4}$ & 0.0026  \\\hline
 \end{tabular}
 \caption{Simulation parameters for the used models.
 \label{tab:Mdotmet1}
 }
\end{table}
}%

\section{$\dot{M}$-discs with constant gas-to-dust ratio}
\label{sec:Mdotdiscconst}

In this section we analyse the disc structure for discs with different $\dot{M}$ rates, but with a constant metallicity of $0.01$.  The different $\dot{M}$ rates correspond to different evolutionary states of the disc's lifetime. We reduce the surface density of the gas, $\Sigma_G$, to realize different states of the evolution, but keep the viscosity at the same value for all simulations. We then analyse the migration behaviour of planetary cores inside the $\dot{M}$ discs by using the torque formula provided by \citet{2011MNRAS.410..293P}.

\subsection{Disc structure}

\begin{figure}
 \centering
 \includegraphics[width=0.9\linwx]{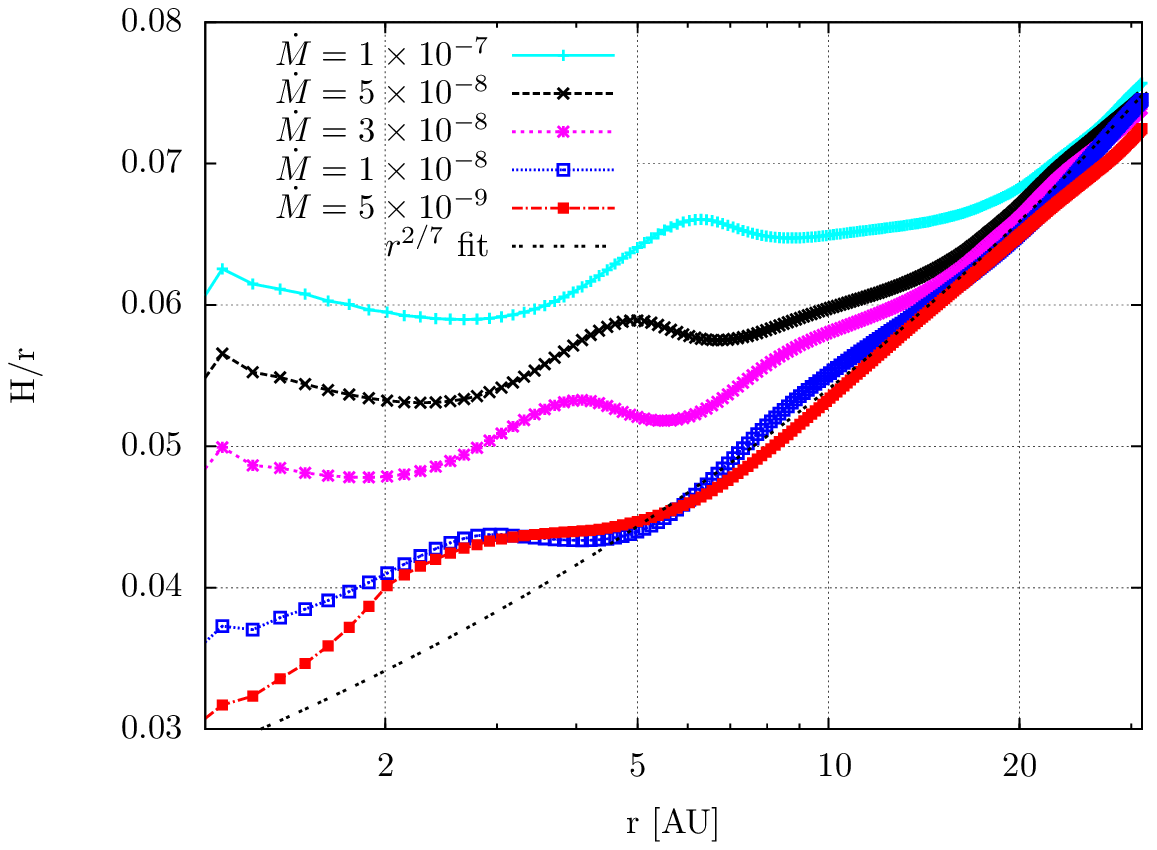}
 \includegraphics[width=0.9\linwx]{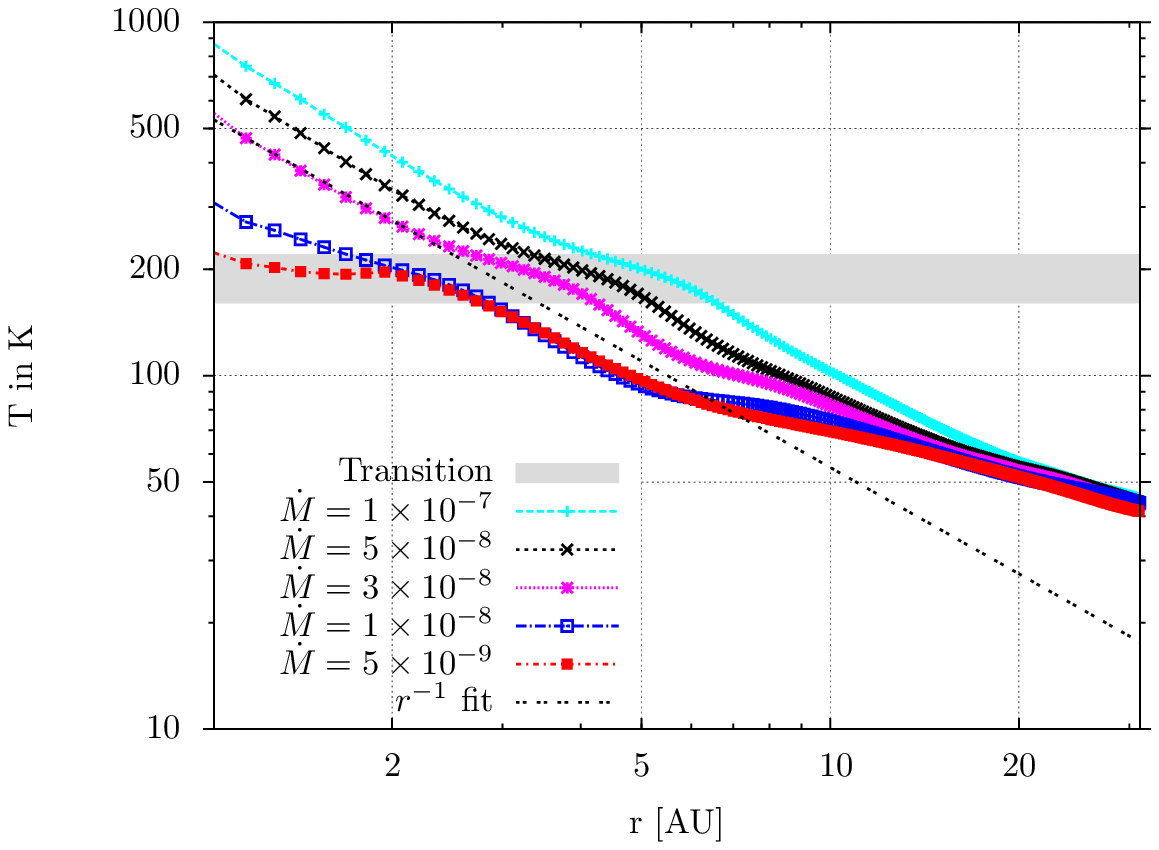}
 \includegraphics[width=0.9\linwx]{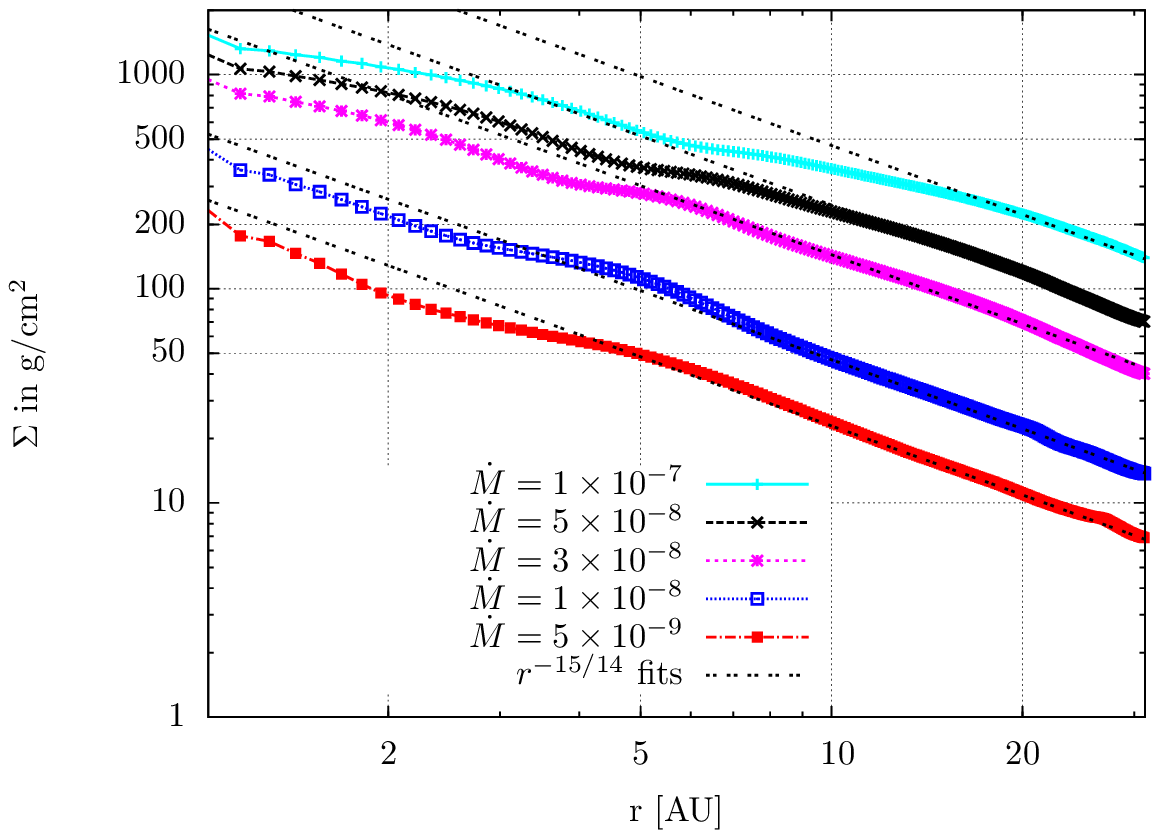}
 \caption{$H/r$-profile (top), temperature (middle), and surface density profile (bottom) of discs with different $\dot{M}$ rates. The profile is taken when the discs have reached their radially constant $\dot{M}$ rate. The black dotted lines represent power-law fits to guide the eyes. The grey area in the temperature plot marks the temperature range where the opacity profile changes due to the melting of ice grains (see Fig.~\ref{fig:kappa}). Please note that we cut the displayed disc at $30$AU to enhance the details in the inner parts of the disc. 
   \label{fig:Hrmdot}
   }
\end{figure}

In Fig.~\ref{fig:Hrmdot}, the $H/r$, temperature, and gas surface density profiles of discs with different $\dot{M}$ are displayed. The $H/r$ profiles show a flaring part in the outer disc and some bumps and dips in the inner part of the disc. These bumps and dips are caused by transitions in the opacity of the disc, for example, at the ice line. This kind of structure was also observed for equilibrium discs with constant viscosity in Paper I. The height of the bumps and the depth of the dips in the disc is decreasing with decreasing $\dot{M}$ rate (which means decreasing $\Sigma_G$). As $\dot{M}$ decreases, so does the viscous heating, hence the temperature and $H/r$ in the inner part of the disc. This also means that the shadowed region of the disc is becoming smaller as the discs density reduces. Likewise, the bumps of the disc structure move towards the central star as $\dot{M}$ decreases. The bumps in the disc come from the opacity transition, which always corresponds to the same temperature region. Therefore less heating in the inner parts of the disc moves these bumps further inside.

Because the gas surface density is reduced more and more, the bumps in the disc become smaller and smaller (as the disc heats less) until they finally disappear completely for low $\dot{M}$. For the $1 \times 10^{-8} M_\odot/yr$ disc there is only a very small bump in the $H/r$ profile visible. This bump vanishes for the $5 \times 10^{-9} M_\odot/yr$ disc, which only shows an increase in $H/r$. In the outer parts of the disc, the structure is very similar for all $\dot{M}$. The heating in the outer disc is dominated by stellar irradiation, which depends on the optical depth ($\Delta \tau = \rho_G \kappa_\star \Delta r$) of the disc. As long as the disc is optically thick ($\Delta \tau > 1$), it can efficiently absorb stellar irradiation and be heated. The disc is then flared with $H/r \propto r^{2/7}$ as found for theoretical calculations in \citet{1997ApJ...490..368C}.

In hydrostatic equilibrium, the temperature is related in midplane to $H/r$ through
\begin{equation}
\label{eq:hydroeq}
 T = \left( \frac{H}{r} \right)^2 \frac{G M_\star}{r} \frac{\mu}{\cal R} \ , 
\end{equation}
where $\mu$ is the mean molecular weight and $\cal R$ the gas constant. The temperature profile (middle in Fig.~\ref{fig:Hrmdot}) therefore reflects the fluctuations in the $H/r$ profile by showing steeper gradients when $H/r$ drops and flatter gradients when $H/r$ increases. These changes are caused by the transition of opacity. The region of change in the opacity is marked in grey in the middle panel of Fig.~\ref{fig:Hrmdot}.

The surface density is reduced by the appropriate factor for changing the $\dot{M}$ rate. In the outer parts of the disc, this can be observed well (bottom in Fig.~\ref{fig:Hrmdot}). However, in the inner parts of the disc this does not seem to be the case exactly. For example, between the $1 \times 10^{-8} M_\odot/yr$ disc and the $5 \times 10^{-8} M_\odot/yr$ disc, the difference in surface density at the inner boundary is only a factor of $\approx2$ instead of $5$. This can be explained by the different viscosity in the inner parts of the disc, since $\nu = \alpha H^2 \Omega$ and $H$ clearly varies for the different $\dot{M}$ values. Lower viscosity caused by a smaller $H$, results in higher surface density because $\dot{M}$ has to be constant, thus explaining the surface density profiles.

Additionally, there are wiggles in the $\Sigma_G$ profile at $\approx 5$AU for all $\dot{M}$ discs. This also can be explained by changes in the viscosity due to changes in $H$ (see top of Fig.~\ref{fig:Hrmdot}), and $\dot{M}$ has to be constant throughout the disc. The changes of the gradients in the surface density $\Sigma_G$ and in the temperature $T$ have important consequences for the migration inside these discs.

\subsection{Migration maps}

To estimate the torque acting on planets embedded in the discs with different $\dot{M}$, we use the formula by \citet{2011MNRAS.410..293P}, which captures the effects of torque saturation in contrast to \citet{2010MNRAS.401.1950P}, where the torques are fully unsaturated. However, the formula by \citet{2011MNRAS.410..293P} might not be accurate for low-mass planets. In fact, \citet{Lega2013} have found a new additional negative torque, which diminishes the total torque acting on embedded low-mass planets. But these differences do not seem to be that great, so in the absence of a formula that would capture this effect and to avoid large scale numerical simulations, we use the formula of \citet{2011MNRAS.410..293P}, which gives a good approximation. The formula captures the torque caused by Lindblad resonances and horseshoe drag on low-mass planets embedded in gaseous discs in the presence of viscous heating and thermal diffusion. The formula does not include stellar heating, but stellar heating only changes the disc structure and not the mechanism responsible for outward migration (Paper I). This formula was also tested against 3D simulations in \citet{2011A&A...536A..77B}, who found a good agreement for $20M_{Earth}$ planets.

\begin{figure}
 \centering
 \includegraphics[width=1.0\linwx]{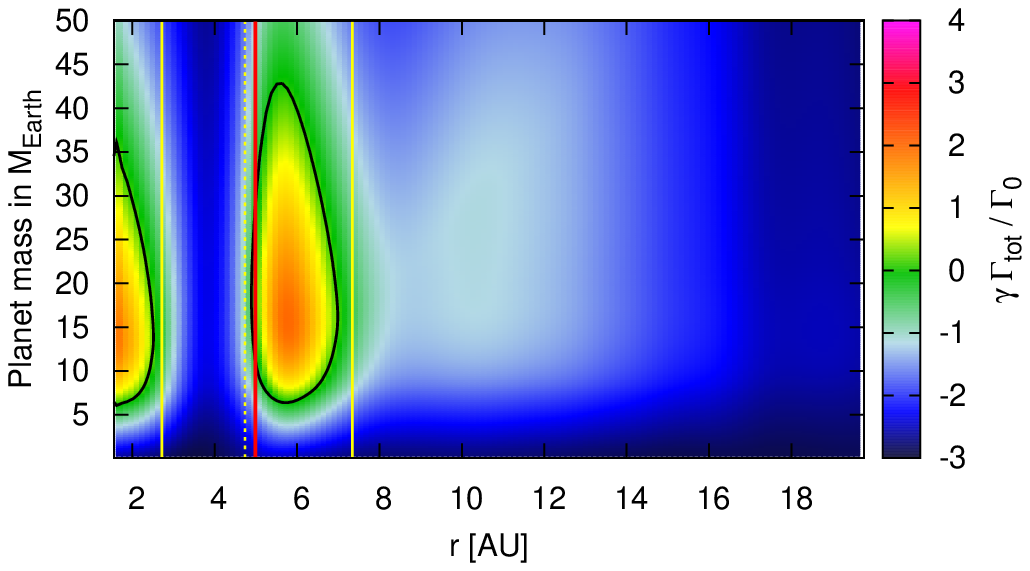}
 \includegraphics[width=1.0\linwx]{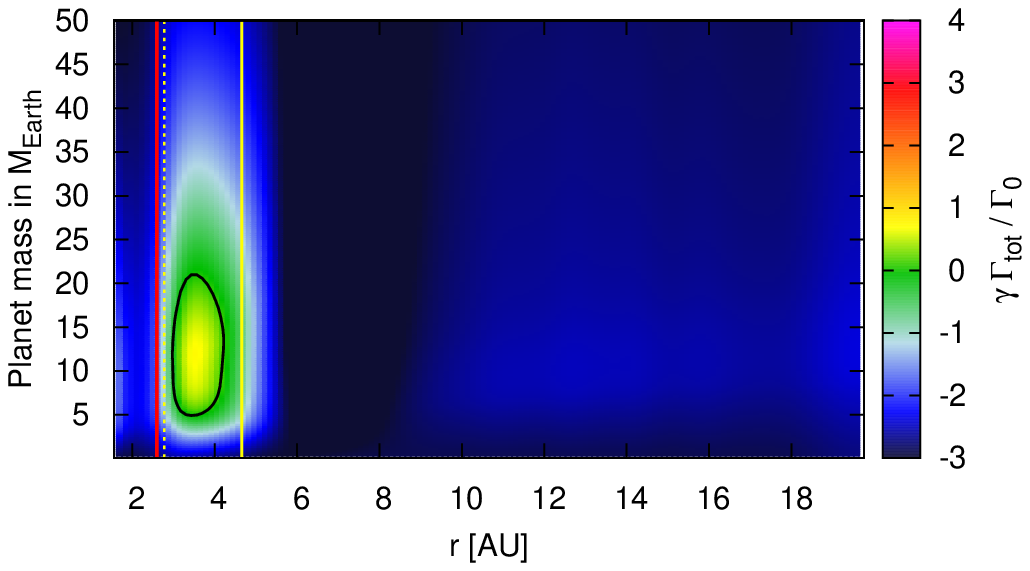}
 \includegraphics[width=1.0\linwx]{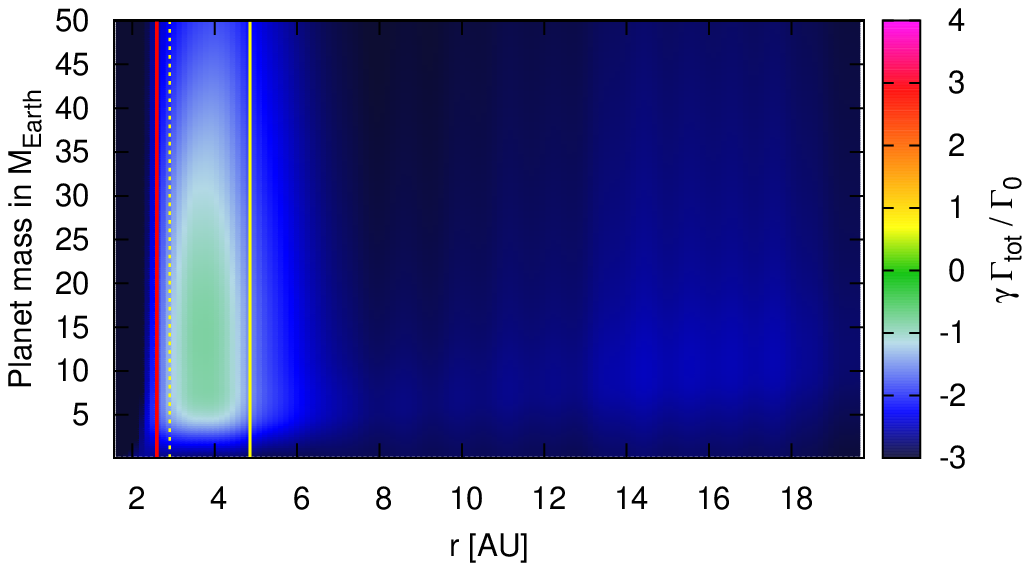}
 \caption{Torque acting on discs with different $\dot{M}$, with $5 \times 10^{-8} M_\odot/yr$ (top), $1 \times 10^{-8} M_\odot/yr$ (middle) and $5 \times 10^{-9} M_\odot/yr$ (bottom) for the applied \citet{2011MNRAS.410..293P} formula. The black lines encircle the regions of outward migration for the \citet{2011MNRAS.410..293P} formula. The vertical solid yellow lines mark the outer edge of the zero-migration region for the unsaturated torques \citep{2010MNRAS.401.1950P}, while the dashed-yellow line marks the inner edge of the zero migration region of the unsaturated torques. The vertical red lines indicate the ice line at $170$K.
   \label{fig:Migmet001}
   }
\end{figure}

The formula of \citet{2011MNRAS.410..293P} is very complex, so we do not explain the whole formula here. The total torque acting on an embedded planet is a composition of its Lindblad torque and its corotation torque:
\begin{equation}
 \Gamma_{tot} = \Gamma_L + \Gamma_C \ .
\end{equation}
The Lindblad torque depends on the gradients of temperature $T \propto r^{-\beta}$ and gas surface density $\Sigma_G \propto r^{-s}$. It is given in \citet{2011MNRAS.410..293P} by
\begin{equation}
\label{eq:Lindblad}
 \gamma \Gamma_L / \Gamma_0 = -2.5 - 1.7 \beta + 0.1 s \quad \mathrm{and} \quad \Gamma_0 = \left(\frac{q}{h}\right)^2 \Sigma_P r_p^4 \Omega_P^2 \ ,
\end{equation}
where $q$ is the mass ratio between planet and star, $\Sigma_P$ the gas surface density of the disc at the planet's location, and $r_P$ the distance of the planet to the host star. One can clearly see that a change in the gradient of temperature influences the Lindblad torque. The same applies to the corotation torque, which strongly depends on the gradient of entropy, $S \propto r^{-\xi}$, with $\xi = \beta - (\gamma - 1.0) s$. The largest contribution of the corotation torque arises from the entropy related horseshoe drag, which is given by
\begin{equation}
\label{eq:horse}
  \gamma \Gamma_{hs,ent} / \Gamma_0 = 7.9 \frac{\xi}{\gamma} \ .
\end{equation}
This indicates that we expect a change in the migration rate, when the temperature changes significantly. This is the case in the inner parts of the discs, where the aspect ratio $H/r$ shows fluctuations (as $H/r$ is proportional to $T$). In fact, in Paper I we found that outward migration only seems possible in regions of the disc where $H/r$ drops. The torque formula that accounts for saturation effects \citep{2011MNRAS.410..293P} will result in a smaller region of outward migration than with the unsaturated torque formula \citep{2010MNRAS.401.1950P}, which can clearly be seen in Fig.~\ref{fig:Migmet001}. Additionally, the formula for the unsaturated torques is independent of the planetary mass, so that outward migration will be found for all planetary masses for the \citet{2010MNRAS.401.1950P} formula, in contrast to the torque formula that accounts for saturation, because torque saturation depends on the planetary mass \citep{2011MNRAS.410..293P}.

Compared to the equilibrium discs with constant viscosity presented in Paper I that had $s_{eq}=0.5$, we now have a much steeper surface density gradient with $s_{\dot{M}} \approx 15/14$. This reduces not only the entropy gradient in the disc and therefore the entropy related horseshoe drag, but also the barotropic part of the horseshoe drag, which is given by
\begin{equation}
\label{eq:baro}
 \gamma \Gamma_{hs,baro} / \Gamma_0 = 1.1 \left( 1.5 - s \right) \ ,
\end{equation}
by $\approx 50\%$ compared to equilibrium discs with $s_{eq}=0.5$. The effect on the Lindblad torque (eq.~\ref{eq:Lindblad}), however, is minimal. The expected total effect would be a reduced region of outward migration in the accreting discs compared to the equilibrium discs.

In Fig.~\ref{fig:Migmet001} we present the migration maps for discs with $\dot{M} = 5 \times 10^{-8} M_\odot/yr$, $\dot{M} = 1 \times 10^{-8} M_\odot/yr$, and $\dot{M} = 5 \times 10^{-9} M_\odot/yr$, for which the disc structures are displayed in Fig.~\ref{fig:Hrmdot}. For the highest $\dot{M}$ disc model, we have two separated regions of outward migration, which are encircled by black lines in the figure. The first region is from the inner edge of the disc up to $\approx 3$AU, the second is from $\approx 5$AU$<r_P<7$AU. These regions compare nicely to regions in the disc where $H/r$ drops. A drop in $H/r$ corresponds to an increase in the temperature gradient resulting in an increased entropy gradient, hence a stronger entropy related corotation torque leading to outward migration. These changes in $H/r$ are caused by opacity transitions, as was stated for equilibrium discs in Paper I. In fact, in the temperature profile shown in Fig.~\ref{fig:Hrmdot} (middle), the grey area marks the region where the opacity changes, which shows a shallower temperature gradient. For the $\dot{M} = 5 \times 10^{-8} M_\odot/yr$ disc, this grey region (from $\approx 3$AU to $\approx 5$AU) corresponds nicely to the region of inward migration shown in Fig.~\ref{fig:Migmet001}.

Compared to the studies of equilibrium discs with constant viscosity (as in Paper I) where the minimum mass for outward migration was $\approx 5M_E$, it is now $\approx 8M_E$. This difference has its origin in the different disc structures. Since the $\dot{M}$ disc features a much steeper surface density gradient $s_{\dot{M}}\approx15/14$ compared to the equilibrium disc $s_{eq}=0.5$, it reduces the different positive contributions of the torque, so that very low-mass planets no longer migrate outwards. For low planetary masses, the horseshoe drag tends towards the linear corotation torque, resulting in inward migration. In fact, for low-mass planets ($<5M_{Earth}$), new studies have shown an additional negative torque \citep{Lega2013}. The maximum mass of outward migration, $\approx 40M_E$, seems to be the same as in Paper I. However, one should be sceptical about the results for high-mass planets, because the formula of \citet{2011MNRAS.410..293P} was derived in the linear regime for low-mass planets and not for large planets that can even start to open up partial gaps.

We do not show the migration map for $\dot{M} = 3 \times 10^{-8} M_\odot/yr$ because it shows just a transition state between $\dot{M} = 5 \times 10^{-8} M_\odot/yr$ and $\dot{M} = 1 \times 10^{-8} M_\odot/yr$, since they have no new features, which can also be guessed from the structures shown Fig.~\ref{fig:Hrmdot}. However, in this case the inner and outer regions of outward migration shrink and are shifted inwards. In fact, when arriving at $\dot{M} = 1 \times 10^{-8} M_\odot/yr$ only one region of outward migration remains in our computational domain. This region is then much narrower in size and valid for a smaller range of planetary masses. But, it also implies that outward migration is possible for planets with slightly lower mass ($\approx 5M_{Earth}$).

The migration map for $\dot{M} = 5 \times 10^{-9} M_\odot/yr$ (bottom in Fig.~\ref{fig:Migmet001}) looks completely different than for the higher $\dot{M}$ models. In fact, there are no more regions of outward migration left. As stated before, outward migration is stronger if $H/r$ drops, but for this case, no drop in $H/r$ can be observed (see Fig.~\ref{fig:Hrmdot}), leading to inward migration for all planetary masses and at all orbital distances. However, in the region where we observed outward migration for higher $\dot{M}$ discs, the inward migration is slowest. In the region between $1.5$ and $2$AU, the inward migration is actually increased compared to the rest of the disc. This is supported by the temperature gradient of the disc, which actually seems to be partly positive (see Fig.~\ref{fig:Hrmdot}), which can result in a negative corotation torque. But, keep in mind here that the actual migration rate is lower for reduced disc masses, since $\Gamma_0$ scales linearly with $\Sigma_G$, see eq.~\ref{eq:Lindblad}. 

Interestingly, only the torque formula that accounts for saturation gives a negative torque for the whole disc, while the fully unsaturated torque formula \citep{2010MNRAS.401.1950P} still predicts outward migration. In fact, in an $\dot{M}$ disc, one can relate the total unsaturated torque, which consists of the Lindblad torque (eq.~\ref{eq:Lindblad}), and the barotropic- (eq.~\ref{eq:baro}) and entropy-related (eq.~\ref{eq:horse}) parts of the corotation torque, to the flaring index of the disc. Assuming that $\Sigma_G \nu$ is independant of $r$, as well as of $\alpha$ (which gives $s=3/2-\beta$), and that $\gamma=1.4$, one finds by using eq.~\ref{eq:hydroeq}
\begin{equation}
 \label{eq:flaringtorque}
 \Gamma_{tot} / \Gamma_0 = 1 - 10 b \ ,
\end{equation}
where $b$ is the flaring index of the disc. This indicates that even the unsaturated torque is negative in discs with flaring indices greater than $0.1$. By interpolating the trend in the $H/r$ profile towards even lower $\dot{M}$ discs, we speculate that the unsaturated torque formula will predict inward migration as $\dot{M}$ becomes lower and lower. However, the torques saturate in time, so that the formula for the unsaturated torques should not be used, especially for low-$\dot{M}$ discs (see discussion in section~\ref{sec:discussion}).

\section{$\dot{M}$-discs with different metallicity}
\label{sec:Mdotdiscvary}

In this section we explore the influence of different metallicities on the disc structure and the corresponding migration maps. A reduced metallicity in small grains could happen in time as the first planetesimals and embryos form in the disc and reduce the amount of heavy elements in the disc. Additionally, radial drift could wash out the dust grains, reducing the metallicity. A higher metallicity in small $\mu m$ size dust grains in the late stages could be achieved by collisional grinding of planetesimals that produce small particles. 

The change in metallicity has many implications, also for planet formation. For example, a higher metallicity helps the streaming instability operate \citep{2007ApJ...662..627J} and planetesimals to form \citep{2007Natur.448.1022J}, while a lower metallicity hinders the streaming instability to work.

\subsection{Higher metallicity}
\label{subsec:highmet}

\begin{figure}
 \centering
 \includegraphics[width=1.0\linwx]{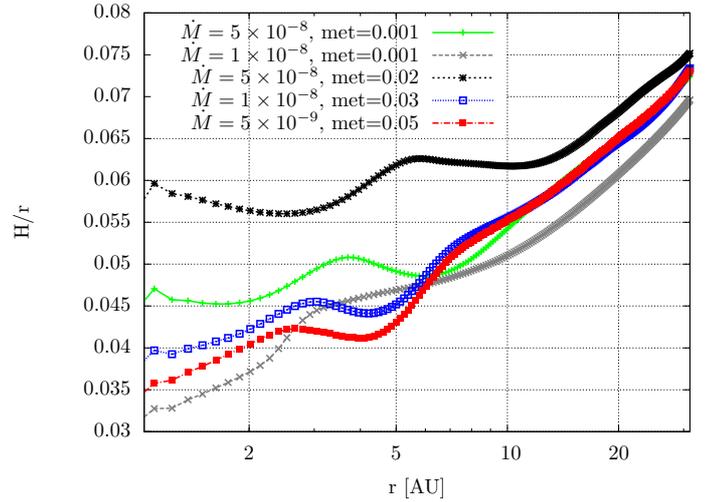}
 \caption{$H/r$-profile of discs with different $\dot{M}$ rates and metallicities. The profile is taken when the discs have reached their radially constant $\dot{M}$ rate. The colour coding of the high-metallicity discs matches the $\dot{M}$ rates displayed in Fig.~\ref{fig:Hrmdot}. Please note that we cut the displayed disc at $30$AU to enhance the visibility in the inner parts of the disc.
   \label{fig:Hrmdot2}
   }
\end{figure}

A change in the metallicity of the disc translates to a change in the opacity of the disc. If the metallicity of a disc is doubled, so is the opacity. If the metallicity increases, the cooling rate of the disc decreases as can be seen from eq.~\ref{eq:raddif}. In Fig.~\ref{fig:Hrmdot2}, it is shown that the aspect ratio $H/r$ of discs increases with increasing metallicities, compared to the low-metallicity cases (Fig.~\ref{fig:Hrmdot}).

In the outer parts of the disc, the structure is quite similar to the one shown for the low-metallicity discs (Fig.~\ref{fig:Hrmdot}). This is not obvious, since the cooling rate reduces with increasing metallicity while the absorption of stellar irradiation increases. However, owing to more grains in the disc, the absorption of stellar irradiation (which is $\propto \rho \kappa_\star$) will happen for lower densities, which implies that it happens at a higher altitude in the disc. Therefore, the heat that transfers down to midplane is actually reduced. But this effect is compensated by the reduced cooling, so the resulting disc structure in the outer parts is the same.

Clearly the aspect ratio is higher in the inner parts of the disc for the $\dot{M} = 5 \times 10^{-8} M_\odot/yr$ disc with a metallicity of $0.02$ compared to the disc with a metallicity of $0.01$ shown in Fig.~\ref{fig:Hrmdot}. We recall that the inner part of the disc is dominated by viscous heating. The viscous heating does not depend on the metallicity, but only on the gas density. The higher metallicity reduces the cooling rate and therefore the balance between viscous heating and cooling turns to higher temperatures, hence to higher $H$. This effect, of course, diminishes as the $\dot{M}$ rate is reduced and therefore viscous heating as well.

Interestingly, not only is the aspect ratio greater for higher metallicity discs, but the bumps in the inner parts of the disc are also more pronounced. For the $\dot{M} = 5 \times 10^{-9} M_\odot/yr$ disc, the bumps are actually visible for the first time. The increase in the bumps inside the disc structure is related to the reduced cooling rate in the high-metallicity discs. This can have important implications on the migration of giant planet cores.

\begin{figure}
 \centering
 \includegraphics[width=1.0\linwx]{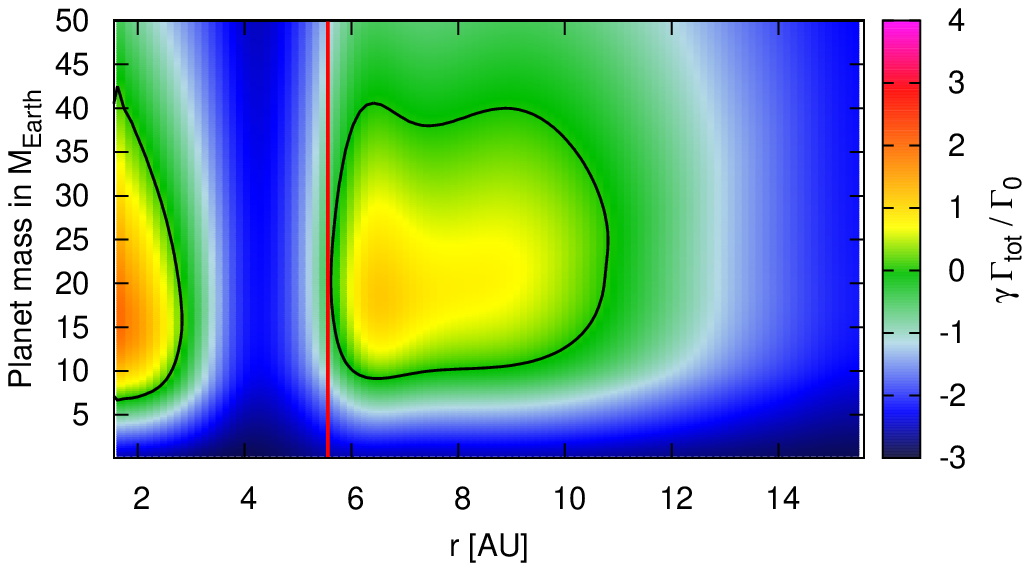}
 \includegraphics[width=1.0\linwx]{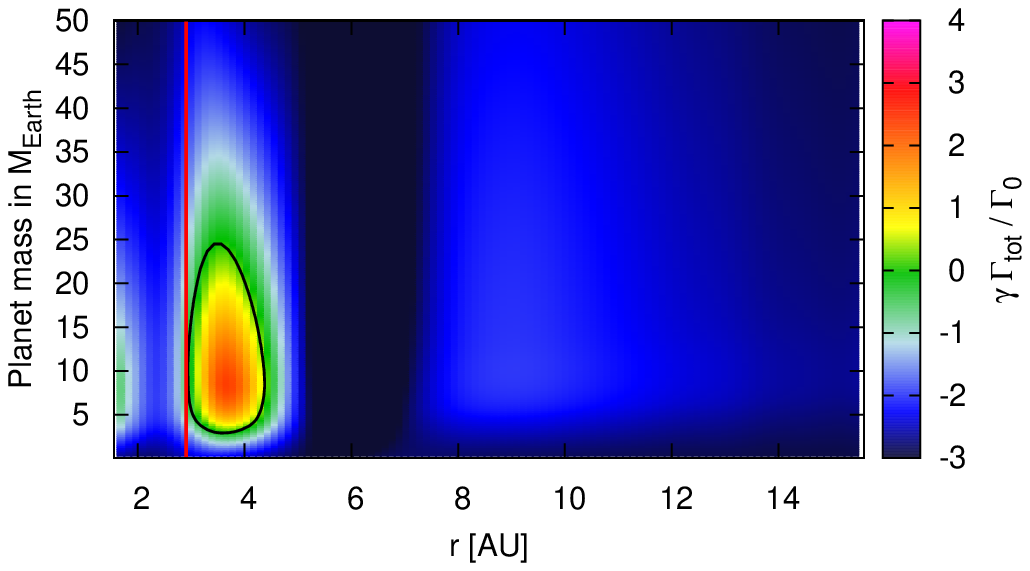}
 \includegraphics[width=1.0\linwx]{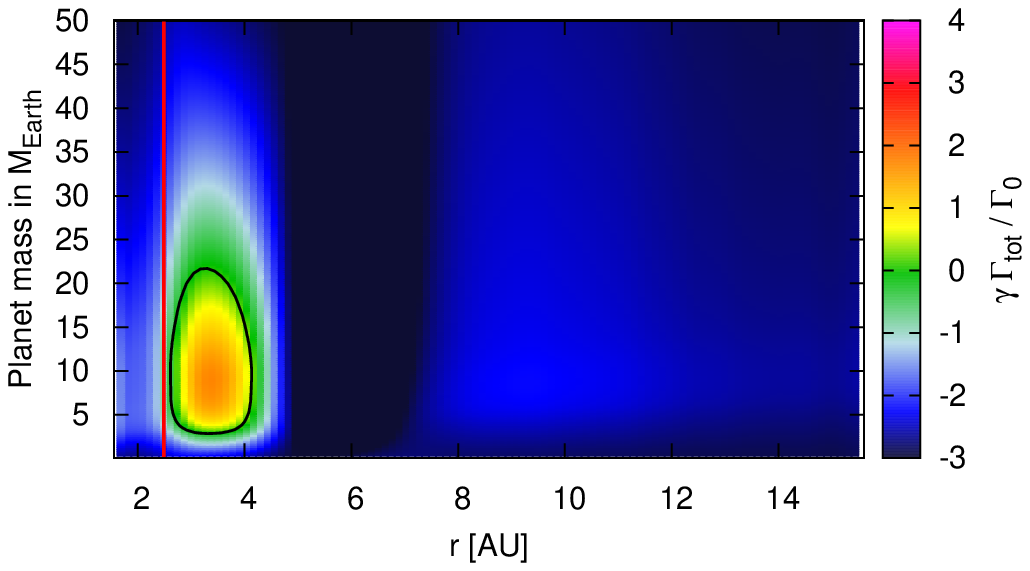}
 \caption{Torque acting on discs with different $\dot{M}$ and metallicity, with $5 \times 10^{-8} M_\odot/yr$ and $0.02$ metallicity (top), $1 \times 10^{-8} M_\odot/yr$ and $0.03$ metallicity (middle), and $5 \times 10^{-9} M_\odot/yr$ and $0.05$ metallicity (bottom). The black lines encircle the regions of outward migration. The vertical red lines indicate the ice line at $170$K. The migration maps correspond to the disc profiles shown in Fig.~\ref{fig:Hrmdot2}.
   \label{fig:Migmet002}
   }
\end{figure}

In Fig.~\ref{fig:Migmet002}, we present the migration maps for the same $\dot{M}$ discs as in Fig.~\ref{fig:Migmet001}, but with different metallicities. The metallicities are $0.02$ (top), $0.03$ (middle), and $0.05$ (bottom). The clear difference for the migration maps displayed here with the ones in Fig.~\ref{fig:Migmet001} is that the regions of outward migration are much larger and more extended. Additionally, the migration map for the $5 \times 10^{-9} M_\odot/yr$ disc shows outward migration, which was not visible for the disc with a metallicity of $0.01$. This is clearly related to the change in the disc structure in the inner parts (see Fig.~\ref{fig:Hrmdot2}), where a bump now appears.

\subsection{Lower metallicity}
\label{subsec:lowmet}

Owing to the formation of planetesimals and planetary embryos, which require dust and ice grains, the amount of small solids can decrease, if there is no additional source of small grains. Discs with reduced metallicity have reduced opacity inside the disc, which works in the same way as described in section \ref{subsec:highmet}. The aspect ratio of discs with a metallicity of $0.001$ are also displayed in Fig.~\ref{fig:Hrmdot2}.

Clearly, lower metallicity leads to a higher cooling rate (eq.~\ref{eq:raddif}), which explains the drop in $H/r$ in the inner parts of the disc. The outer parts of the disc remain flared and at about a constant level, as explained in section~\ref{subsec:highmet}. For the $\dot{M} = 1 \times 10^{-8} M_\odot/yr$ disc, there is a dramatic change in the disc structure because no bumps in the discs profile are visible any more, which are still visible in the case of a metallicity of $0.01$.

\begin{figure}
 \centering
 \includegraphics[width=1.0\linwx]{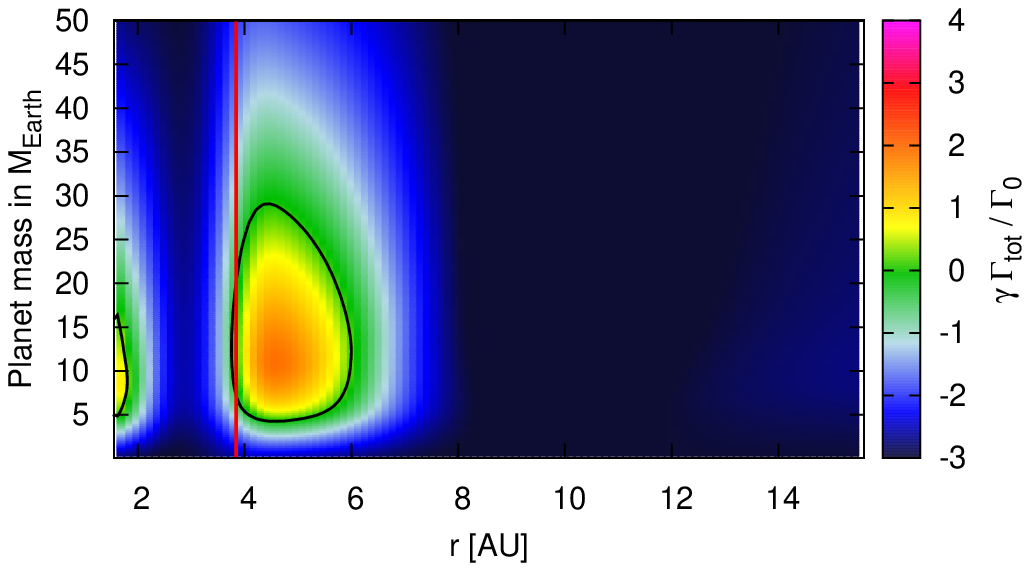}
 \includegraphics[width=1.0\linwx]{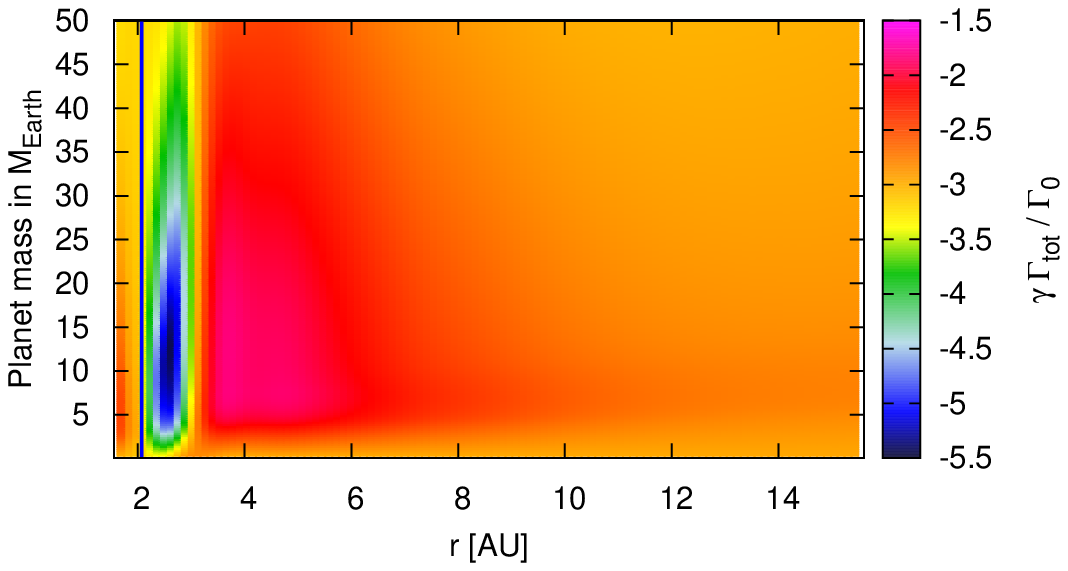}
 \caption{Torque acting on discs with different $\dot{M}$ and metallicity, with $5 \times 10^{-8} M_\odot/yr$ and $0.001$ metallicity (top) and $1 \times 10^{-8} M_\odot/yr$ and $0.001$ metallicity (bottom). The black lines encircle the regions of outward migration. The vertical red (top) and blue (bottom) lines indicate the ice line at $170$K. The migration maps correspond to the disc profiles shown in Fig.~\ref{fig:Hrmdot2}. There is a different colour scale for the bottom plot.
   \label{fig:Migmet0001}
   }
\end{figure}

In Fig.~\ref{fig:Migmet0001} the migration maps for the low-metallicity discs are displayed. In the $\dot{M} = 5 \times 10^{-8} M_\odot/yr$ case (Fig.~\ref{fig:Migmet0001}, top), the regions of outward migration are much smaller than in the $0.01$ metallicity case (top of Fig.~\ref{fig:Migmet001}). In fact, the inner part region of outward migration has now completely disappeared, and the outer region is only valid for much lower masses. Additionally, the outer region of outward migration is shifted inwards from $\approx 5$AU to $\approx 3.8$AU, which can be explained by the higher cooling rate in the disc, which shifts the ice line further inwards.

In the case of $\dot{M} = 1 \times 10^{-8} M_\odot/yr$ (Fig.~\ref{fig:Migmet0001}, bottom), the changes are dramatic because no region of outward migration exists any more. In fact, the region inside the ice line ($r>r_{ice}$) now features a region of enhanced inward migration. In all other displayed migration maps, we observe a region of outward migration (or at least with slower inward migration). At this point in the disc, the aspect-ratio profile undergoes a steep gradient (Fig.~\ref{fig:Hrmdot2}), which actually refers to a positive gradient in temperature, which in turn leads to a negative entropy-related corotation torque, resulting in an even stronger negative total torque. The reason for that is probably related to the transition of viscous heating to stellar heating. Since viscous heating is operating in the inner (midplane) parts of the disc, heat is produced there. Stellar irradiation heats the upper layers of the disc, which then transport heat down towards midplane. At $\approx 2.5$AU, stellar heating is getting absorbed effectively, which heats the midplane, but viscous heating already seems to diminish at a shorter distance to the star. Therefore a higher temperature can be observed at $\approx 2.5$AU compared to the inner parts of the disc, which leads to a positive temperature gradient in the disc.

\section{Influence of the interchange between $\Sigma_G$ and $\nu$}
\label{sec:InfluenceSnu}

Because the mass flux $\dot{M}$ through a disc is proportional to the gas surface density $\Sigma_G$ and the viscosity $\nu$, a change in $\Sigma_G$ could be compensated by a change in $\nu$ to still have the same $\dot{M}$. We now want to construct a scenario where $\dot{M}$ and the thermal structure are the same, but where $\Sigma_G$ and $\nu$ are not. For changing $\nu$, we change $\alpha$.

We now look at the different heating and cooling sources of the disc. The viscous heating, which is the dominant heat source in the inner parts of the disc, is proportional to $\Sigma_G \nu$ ($\dot{M}$):
\begin{equation}
\label{eq:Qplus}
  Q^+ = \frac{9}{8} \Sigma_G \nu \Omega_K^2 2 \pi r \delta r \ ,
\end{equation}
which implies that an interchange between $\Sigma_G$ and $\nu$ would not change the viscous heating. The radiative cooling is given by 
\begin {equation}
\label{eq:raddif}
 \qquad \vec{F}  =  -  \frac{\lambda c}{\rho_G \kappa_R} \, \nabla E_R\ ,
\end{equation}
where $c$ is the speed of light, $\lambda$ the flux-limiter \citep{1989A&A...208...98K}, and $E_R$ the radiation energy. To get the same cooling rate, a change in $\rho_G$ has to be balanced by a change in $\kappa_R$. The same applies to the absorption of stellar photons, which scales with $\rho_G \kappa_\star$. If the gas density is increased by a factor $s_f$, the opacity, hence the metallicity, has to decrease by a factor $s_f$. The ratio of $\Sigma_G / \Sigma_Z$ will change, but this will not change the thermal structure of the disc. This actually means that the surface density of heavy elements $\Sigma_Z$ stays constant. 

As a result, we can use the data of the discs presented in the previous section and scale them by scaling factor $s_f$ to exchange $\Sigma_G$ and $\nu$. When increasing $\nu$ by $s_f$, the gas surface density $\Sigma_G$ has to decrease by $s_f$. At the same time, $\Sigma_Z$ has to be constant, but this means that the metallicity of the disc increases, so in scenario like that, the opacities $\kappa_R$, $\kappa_P$, and $\kappa_\star$ increase by $s_f$ as well.  

\begin{figure}
 \centering
 \includegraphics[width=1.0\linwx]{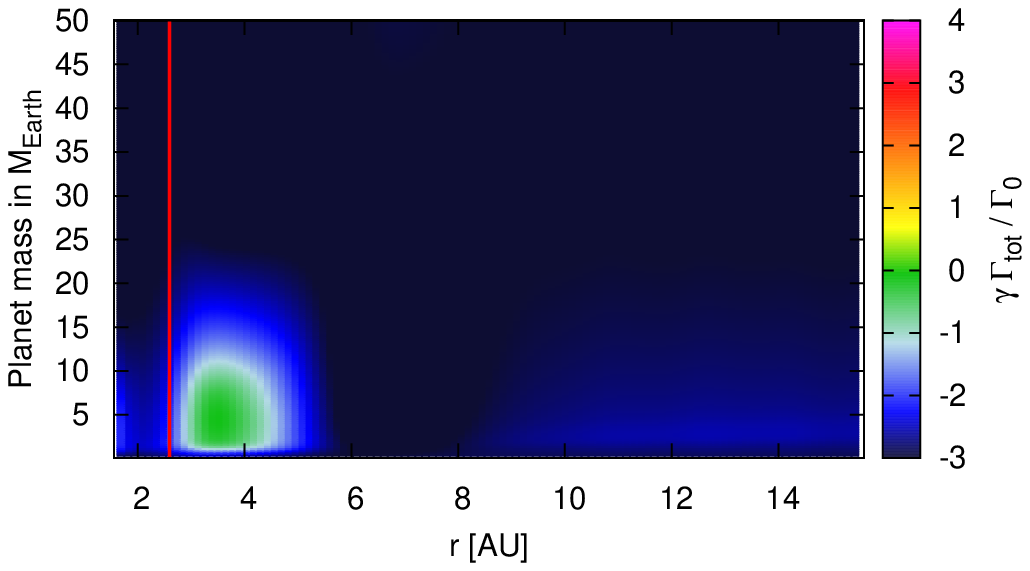}
 \includegraphics[width=1.0\linwx]{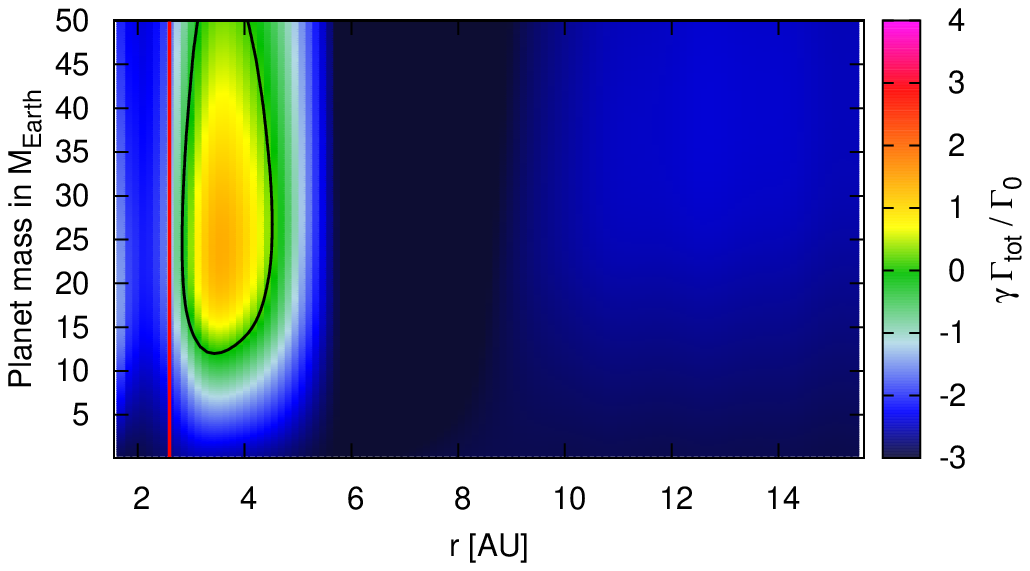}
 \caption{Torque acting on discs with $\dot{M} = 1 \times 10^{-8} M_\odot/yr$ for a rescaling of $\Sigma_G \nu$ by a factor of $s_f=0.5$ (top) and a factor of $s_f=2$ (bottom). This means that the disc displayed on top has a metallicity of $0.005$, while the disc on the bottom has a metallicity of $0.02$. The black lines encircle the regions of outward migration. The vertical red lines indicate the ice line at $170$K.
   \label{fig:Migscale}
   }
\end{figure}

In Fig.~\ref{fig:Migscale} we present the migration maps of an $\dot{M} = 1 \times 10^{-8} M_\odot/yr$ disc, for two different cases for exchanging $\Sigma_G$ and $\nu$. At the top, the viscosity is decreased by a factor of $2$, and in the bottom plot it is increased by a factor of $2$. This means that the top disc has $0.005$ metallicity and the bottom one $0.02$. 

These two migration maps significantly differ from each other and from the one presented in fig.~\ref{fig:Migmet001}. In the case of low viscosity (top in Fig.~\ref{fig:Migscale}), no outward migration is detected any more. But in the case of high viscosity, a much larger region of outward migration is visible. However, outward migration in this case starts at much higher planetary masses ($12 M_{Earth}$) than in the normal viscosity case presented in fig.~\ref{fig:Migmet001}, where outward migration starts for $\approx 5M_{Earth}$.

The reason for this is the scaling of the thermal diffusion coefficient $\chi$ in the formula of \citet{2011MNRAS.410..293P}, which is not linear in $\rho_G \kappa_R$. It reads as
\begin{equation}
 \chi = \frac{16 \gamma (\gamma - 1) \sigma T^4}{3 \kappa_R \rho_G^2 H^2 \Omega_P^2} \ ,
\end{equation}
where $\sigma$ is the Boltzmann constant. A factor of $4$ is missing in \citet{2011MNRAS.410..293P} according to \citet{2011A&A...536A..77B}. Additionally, the saturation parameters for the corotation torque and horseshoe drag also depends on the viscosity of the disc. These saturation parameters are responsible for the transition from horseshoe drag towards the linear corotation torque, which is much less than the horseshoe drag. This can then lead to inward migration, changing the migration maps.

This non-linearity causes different migration behaviour for discs with different $\nu$ to $\Sigma_G$ ratios, even if they feature the same $\dot{M}$ value and the same thermal structure. In total, one could roughly say that lower viscosity allows outward migration for lower planetary masses (unless it gets too low and no outward migration is possible any more), while a higher viscosity allows outward migration for higher planetary masses, where the minimum mass required for outward migration is also increasing. This is crucial for N-Body simulations of planetary embryos in evolving discs, where just stating an $\dot{M}$ value would not be enough.

\section{Consequences for planet formation}
\label{sec:discussion}

The presented migration maps have important consequences for the resulting structures of planetary systems. \citet{2010ApJ...715:L68} stated that planets formed in a disc, would migrate and then sit at the zero-migration radius, which is independent of the planetary mass in their case, because they use the torque formula of \citet{2010MNRAS.401.1950P} that only accounts for the unsaturated torques. This picture is contradicted by the results presented here, since we use the torque formula with saturation \citep{2011MNRAS.410..293P}. 

The planets in \citet{2010ApJ...715:L68} consequently move inwards with the zero-torque radius in time as the disc disappears, which can also be observed from the migration maps presented here. However, they used a simple 1D model for their disc evolution and did not account for stellar heating, which clearly changes the disc structure and the migration pattern compared to fully radiative discs (Paper I). In particular, eq.~\ref{eq:flaringtorque} shows that the flaring index should be determined accurately for a good estimate of the unsaturated torque, but non-stellar irradiated discs cannot be flared in the outer parts (Paper I). The main difference between our model and the \citet{2010ApJ...715:L68} model originate in the more complex disc structure presented here and from the different torque formula that accounts for saturation effects. 

\citet{2010ApJ...715:L68} also stats that in the late evolutionary stages of the disc, starting at $\approx 4.0$Myr, the surface density is so low that it cannot transfer enough angular momentum to the planet for its orbital radius to evolve as fast as the equilibrium radius. This means that the planets detach from the evolution of the disc and are left behind. The beginning of their late stage of $\approx 4.0$Myr matches the surface density at $1$AU quite well in our $\dot{M} = 5 \times 10^{-9} M_\odot/yr$ model with a metallicity of $0.01$, where we no longer observe a zero-torque radius any more for the torques that are prone to saturation \citep{2011MNRAS.410..293P}. This would indicate that planetary cores could start inward migration while the disc is disappearing and cannot safely wait and detach themselves from the disc as proposed by \citet{2010ApJ...715:L68}. Therefore this could make the survival of planets in the outer parts of the disc in the late stages much harder. This problem becomes even more severe if the metallicity of the disc drops in time (e.g. due to the formation of planetesimals out of dust grains) as can be seen in Fig.~\ref{fig:Migmet0001}, where only inward migration is observed for $\dot{M} = 1 \times 10^{-8} M_\odot/yr$.

The results of \citet{2010ApJ...715:L68} could be interpreted in such a way that the evolution of the disc naturally would account for the occurrence of Super-Earth and Mini-Neptune planets in the inner parts of the solar system. However, if the metallicity of the disc is high in the late stages of the evolution, the zero-migration radius for these planets is at $\approx 4$AU (see Fig.~\ref{fig:Migmet002}), which is much larger than in the model of \citet{2010ApJ...715:L68}, even though we used the torque formula when accounting for saturation here. Additionally, in the low-metallicity case, the zero-torque radius for $\dot{M} = 5 \times 10^{-9} M_\odot/yr$ using the fully unsaturated torque is much further out in our simulations ($\approx5$AU) compared to $\approx 2$AU in \citet{2010ApJ...715:L68}, which is clearly a consequence of the different disc models.

From our simulations, we propose two different ways to form hot Neptunes. Either they form in the outer disc, where they get trapped in the outer region of outward migration and are then released at a late stage where they migrate inwards to the inner edge of the disc (low-metallicity scenario). Or they assemble in the inner part of the disc from migrating embryos that were too small to be trapped in the regions of outward migration. The second scenario can happen for low and high metallicities, since both scenarios require a minimal planetary mass to halt inward migration.

The migration maps for a metallicity of $0.01$ (Fig.~\ref{fig:Migmet001}) can also be used as a guideline to discuss the evolution of giant planets during their formation. The core of a giant planet most likely forms inside the ice line ($r>r_{ice}$) at the early stages of the disc. The reason for the core to actually grow to become a giant planet and not stay a hot Neptune might be found in the different formation time of the core itself. A faster formation of the core provides more time to accrete gas to become a gas giant than a later formation of the core where the planet is stuck at Neptune size. We assume that a core is trapped at the outer zero-migration radius where it can grow until it reaches a mass of $\approx 40M_{Earth}$. Thus, at that stage the planet is in a region between $\approx 4$ and $\approx 5$AU, depending on which evolutionary stage the disc is in (the lower $\dot{M}$, the closer to the star the equilibrium radius). When the $\approx 40M_{Earth}$ mass is exceeded, the planet is released and starts migrating towards the central star. A planet of that size grows rapidly due to gas accretion \citep{1996Icar..124...62P} and it starts to open a significant gap inside the disc changing its migration regime to that of type-II-migration. Owing to this migration phase, the giant planet, once fully formed, is significantly closer to the star than the original location of its core, i.e. closer than $\sim 4$AU. The formation of an inner cavity in the disc by photo-evaporation can eventually stop the inward migration of the planet, as suggested by \citet{2012MNRAS.422L..82A}, preventing it from becoming a hot Jupiter.

However, in the case of our own Solar System, this is not sufficient. A mechanism is needed to move the giant planet (Jupiter) outwards and bring it to $5$AU. This consideration supports the Grand Tack scenario \citep{2011Natur.475..206W}, in which the formation of Saturn eventually forces Jupiter to reverse its migration direction and move outwards. In the Grand Tack scenario, Saturn, once formed, migrates faster than Jupiter and catched it in resonance inside their common gap, which causes the migration reversal \citep{2001MNRAS.320L..55M, 2007Icar..191..158M, 2008A&A...482..333P}. A potential problem that is often invoked with this scenario is that a priori Saturn and Jupiter should have followed exactly the same growth-migration histories, so that Saturn could never catch up with Jupiter. Our work now shows that this assumption is actually not true. If Saturn forms after Jupiter, $\dot{M}$ will have decreased in the disc, and the migration map will differ. In fact, as times goes by, the maximum mass for being still trapped in a zero-torque zone decreases, so Saturn should start migration inwards at a lower mass than Jupiter did. Possibly, Jupiter never had inward type-I-migration, while Saturn had, and may even have experienced fast type-III-migration, which leads Saturn to catch up with Jupiter.

Of course this depends on the ratio between the migration speed of the giant planets and the actual accretional evolution of the disc. In addition, opening of a gap by Jupiter would change the thermal structure of the disc, hence changing the migration map for Saturn. A detailed study of this process is beyond the scope of this paper, but it is worth noting that our results qualitatively support the concept of the Grand Tack scenario.

It also seems that it is easier to keep the first cores in discs with higher metallicity, since the minimum mass needed for outward migration is reduced compared to lower metallicity discs. This would potentially make it easier to keep the first cores that form giant planets. Because more cores can potentially be kept inside the disc, it is more likely that these cores can form giant planets. Therefore, high-metallicity discs would imply a higher probability of forming gas giant planets, which is supported by observations.

\citet{2011MNRAS.417.1236H} state that in the ice line and dead zone, the heat transition between viscous and stellar heating can act as a planet trap. In a disc, stellar heating is dominant in the outer parts of the disc and therefore responsible for the flaring part of the disc (see \citet{1997ApJ...490..368C}, Paper I). In the inner parts, viscous heating is dominant. The different heat sources result in a change of the gradient of temperature between these two different regions, which can lead to a region of outward migration, according to \citet{2011MNRAS.417.1236H}. Additionally, they stated that the ice-line (located at $r_{ice}$) can function as a planet trap.

However, in the simulations presented here, we see a different picture. The ice line functions as a region of divergent migration, meaning that planets outside the ice line ($r<r_{ice}$) migrate inwards, while planets located inside the ice line ($r>r_{ice}$) migrate outwards. Additionally, one could argue that the disc is locally radial isothermal at the ice line, because melting and resublimating of ice grains keep the temperature constant for a certain distance around the actual ice line \citep{2010ApJ...715:L68, 2011MNRAS.417.1236H}, which is not explicitly taken into account in our simulations, but it is slightly reflected by the smoothed transition between the different branches of opacity (Fig.~\ref{fig:kappa}). Since the disc is locally radial isothermal, no temperature gradient exists around the ice line, so that outward migration is not possible there, and therefore the ice line cannot act as a point of convergent migration.

The outward migration inside the ice line then stops in a region of the disc where $H/r$ reaches a local minimum, which is a convergent migration radius in the disc. A minimum of $H/r$ can only exist, if $H/r$ increases again, which, in this case, is achieved by stellar irradiation. In that sense, a change form inward to outward migration is caused by stellar irradiation, because it flares the disc. But at the same time the change of the opacity at the ice line, provides a change in the disc structure as the cooling properties of the disc change. In this sense, the divergent migration point is related to the opacity transition at the ice line, and the stopping point of outward migration is related to stellar heating, because the disc becomes flared (Paper I).

The turbulence inside the disc is thought to be driven by the MRI, which would then also account for a non-turbulent inner part of the disc, the so-called dead zone. This zone is basically a region of low viscosity inside the disc, which can have important consequences on the disc structure and on the migration of embedded planets. The study of the influence of the dead zone on disc structures is planned in future work. Along with to that, we plan to use the presented code in 3D to test the influence of embedded planets on the disc structure of stellar irradiated discs and that the migration behaviour is still similar to the one predicted by the torque formulae. With the presented code, these simulations seem to be possible in a reasonable amount of computation time (several weeks).

\section{Summary}
\label{sec:summary}

We presented here hydrodynamical simulations of stellar irradiated accretion discs, featuring viscous heating, radiative cooling, and stellar irradiation. These discs feature different radially constant accretion fluxes $\dot{M}=3\pi \nu \Sigma_G$ through them that represent different stages of the lifetime of an accretion discs, since the $\dot{M}$ rate is reducing as the disc becomes older. We especially focussed on the thermal structure of the disc. From the resulting disc structure, we computed the torque acting on embedded planets by using a torque formula \citep{2011MNRAS.410..293P} to find sweet spots for planetary growth.

At the ice line, the opacity profile (Fig.~\ref{fig:kappa}) changes due to the melting and condensation of ice grains. This transition in opacity changes the stellar heating and radiative cooling rates, resulting in bumps and dips in the inner disc. These bumps and dips in the aspect ratio $H/r$ of the disc translate into changes in viscosity, $\nu = \alpha H^2 \Omega_K$. This causes changes in the gas surface density $\Sigma_G$ because the product of viscosity and surface density is constant (the $\dot{M}$-rate). The resulting disc structure (density and temperature) is therefore not uniform and does not follow a simple power law. Additionally, the bumps and dips are smaller and closer to the central star when the $\dot{M}$ rate is reduced. In the final stages, the disc is flared with no bumps in the inner part.

The thermal structure of the disc depends on the gas surface density $\Sigma_G$ and the viscosity $\nu$ in such away that they are interchangeable, as long as the surface density of heavy elements $\Sigma_Z$ is constant. In other words, for discs with a given $\Sigma_G \times \nu$ product, the same disc structure is reached as long as $\Sigma_Z$ is the same. Additionally, we pointed out that a non linear scaling in the torque calculations exists, which indicates that an interchange between $\Sigma_G$ and $\nu$ leads to a different migration behaviour, even when the thermal disc structure is the same. In fact, a lower disc viscosity results in a possible outward migration for lower mass planets than a higher viscosity disc. We can therefore conclude that the thermal structure is a two-parameter space in $\Sigma_G \nu$ and $\Sigma_Z$, while the migration is a three-parameter space in $\Sigma_G \nu$, $\nu$, and $\Sigma_Z$.

As low-mass cores can migrate, collide, and form bigger objects at convergent migration distances \citep{2013A&A...558A.105P}, the location of these is of high importance. We calculated the migration by applying the \citet{2011MNRAS.410..293P} torque formula to our disc structures. We should point out here that in all the disc scenarios, a minimum planetary mass is needed to find outward migration. This mass lies (dependent on the underlying disc structure) in the range of $\approx 3$ to $\approx 5M_{Earth}$. This is caused by a transition from the horseshoe drag towards the linear corotation torque in viscous discs, which reduces the total torque acting on an embedded planet.

We find a nice match between the regions in the disc where the opacity law changes (and therefore the disc structure) and where the migration changes. In fact, we confirm here the results of Paper I, where we stated that outward migration is more likely in regions where $H/r$ drops. Of course, this means that for smaller bumps and dips in the $H/r$ profile, a smaller region of outward migration is the consequence. For a metallicity of $0.01$ outward migration is no longer supported in the late stages of the disc (low $\dot{M}$).

As the disc accretes onto the star, the metallicity of the disc can change. Our simulations show that a disc with metallicity of $0.001$ actually gives rise to only inward migration, compared to discs with a metallicity of $0.01$. Higher metallicity, on the other hand, allows outward migration for discs with an even lower $\dot{M}$ rate. If the metallicity of the disc increases in the late stages, it is therefore easier to keep planets in the disc with gas-driven migration.

The migration map is therefore not fixed in time, but evolves as $\dot{M}$ and the metallicity change. This is crucial for the migration scenarios. This influences the formation of hot Neptunes that can halt their migration in these planet traps, because it allows for two formation scenarios. Either they form in the outer disc, get trapped at the convergent migration radius, and are then released towards the inner disc in the late stages, or they assemble from smaller bodies that migrate towards the inner disc, which then grow and stay there. Along with that, the evolution of the migration maps also give indications about the final location of giant planets. Especially in the Solar System case our migration maps indicate that Jupiter would be formed well inside $4$AU and thus needs a mechanism that transports it outwards again, which is provided by the Grand Tack scenario \citep{2011Natur.475..206W}.

\begin{acknowledgements}

B. Bitsch and A. Morbidelli have been sponsored through the Helmholtz Alliance {\it Planetary Evolution and Life}. The Nice group is thankful to ANR for supporting the MOJO project. The computations were done on the “Mesocentre SIGAMM” machine, hosted by the Observatoire de la C\^{o}te d'Azur.
We thank W. Kley and C. Dullemond for useful discussions. Additionally, we want to thank an anonymous referee for comments that helped to improve the manuscript.

\end{acknowledgements}

\appendix
\section{Boundary conditions}
\label{ap:boundary}

\subsection{Inner boundaries}

We impose open inner boundaries, which means the velocity can be direct inwards towards the star and material can leave the disc. For the radial velocity at the boundary between active and ghost domain follows:
\begin{equation}
v_r (z) =  \left\{
    \begin{array}{cc} 
    0
     \quad &  \mbox{for} \quad v_{r,A}(z)>0  \\
   v_r (z)  \quad & \mbox{for} \quad  v_{r,A} (z)<0 
    \end{array}
    \right. \ ,
\end{equation}
where $v_{r,A}$ is the radial velocity in the first active grid cell. A negative velocity implies an inward flux (towards the central star). The polar velocity $v_\theta$ is copied from the active domain, and the azimuthal velocity is set via the following equation:
\begin{equation}
 v_{\phi, G} = v_{\phi, A} \sqrt{r_A / r_G} \ ,
\end{equation}
where $r_A$ and $r_G$ are the distances of the active and ghost cells to the central star. Note that $r_A > r_G$.

The density in the ghost cells is copied from the active cells to the ghost cells to mimic the outflow condition. This applies to all other cell-centred quantities ($\epsilon$, $T$, $E_R$) as well.

\subsection{Outer boundaries}

From the starting configuration with a given $H$, the disc evolves into its final state with a different $H$. A change in $H$ also implies a change in the viscosity, $\nu = \alpha c_s H$. But the viscosity is vertically constant, as is predicted by MHD simulations \citep{2012MNRAS.420.2419F}. We impose the same $H/r$ in the ghost and active cell,
\begin{eqnarray}
  H_G/r_G &= &H_A/r_A \nonumber \ , \\
  c_{s,G} &= &c_{s,A} \frac{\Omega_G r_G}{\Omega_A r_A} = c_{s,A} \sqrt{r_A / r_G} \nonumber \ ,
\end{eqnarray}
where we can now calculate the height in the ghost cell $H_G$ and the sound speed $c_{s,G}$. The sound speed in midplane is used to calculate the viscosity in the ghost cells, where an averaged viscosity between active and ghost cells then gives the velocity (since the velocity is defined on the edges of the grid cells):
\begin{equation}
 v_r (z) = -\frac{3 \nu }{2 r} = -\frac{3 \alpha c_s^2 / \Omega}{2 r} \ ,
\end{equation}
where $c_s$ is the sound speed in the ghost cells at midplane height. The minus sign defines an inward (to the star) velocity.

The polar velocity $v_\theta$ is copied from the active domain and the azimuthal velocity is set via the following equation:
\begin{equation}
 v_{\phi, G} = v_{\phi, A} \sqrt{r_A / r_G} \ ,
\end{equation}
where $r_A$ and $r_G$ are the distances of the active and ghost cells to the central star. Note that $r_A < r_G$.

The density in the ghost cells has to be provided by $\dot{M}$. Taking into account that $\rho$ varies with height, we get
\begin{equation}
\label{eq:Mdotz}
 \dot{M} = 3 \pi \nu \Sigma_G = 3 \pi \nu \int_0^z \rho (z') dz' \ .
\end{equation}
The vertical density distribution (hydrostatic equilibrium) is given by
\begin{equation}
\label{eq:rhoz}
 \rho (z) = \rho_0 e^{-z^2 / (2H^2)} \ ,
\end{equation}
with $H=c_s \Omega$ and $\rho_0$ can be determined by putting eq.~\ref{eq:rhoz} in eq.~\ref{eq:Mdotz} and solving as a function of $\dot{M}$. The resulting $\rho_0$ will then be used to calculate $\rho (z)$ in the ghost cells. This way, the density can change in time as the viscosity changes until the equilibrium state is reached, but the disc will always have the same $\dot{M}$ inflow. The other cell-centred quantities are copied from the active ring, as done for the inner boundaries.

\bibliographystyle{aa}
\bibliography{Stellar}
\end{document}